\documentclass[aps,pra,longbibliography,reprint, superscriptaddress]{revtex4-1}

\usepackage{graphicx}
\usepackage{dcolumn}
\usepackage{bm}

\usepackage{array}
\newcommand{\PreserveBackslash}[1]{\let\temp=\\#1\let\\=\temp}
\newcolumntype{C}[1]{>{\PreserveBackslash\centering}p{#1}}
\newcolumntype{R}[1]{>{\PreserveBackslash\raggedleft}p{#1}}
\newcolumntype{L}[1]{>{\PreserveBackslash\raggedright}p{#1}}

\begin{document}

\preprint{APS/123-QED}

\title{Compensation-Free High-Capacity Free-Space Optical Communication Using Turbulence-Resilient Vector Beams}


\author{Ziyi~Zhu}
\affiliation{Department of Physics, University of South Florida, Tampa, FL 33620, USA}

\author{Molly~Janasik}
\affiliation{Department of Physics, University of South Florida, Tampa, FL 33620, USA}

\author{Alexander~Fyffe}
\affiliation{Department of Physics, University of South Florida, Tampa, FL 33620, USA}

\author{Darrick~Hay}
\affiliation{Department of Physics, University of South Florida, Tampa, FL 33620, USA}

\author{Yiyu~Zhou}
\affiliation{The Institute of Optics, University of Rochester, Rochester, New York 14627, USA}

\author{Brian~Kantor}
\affiliation{Department of Physics, University of South Florida, Tampa, FL 33620, USA}

\author{Taylor~Winder}
\affiliation{Department of Physics, University of South Florida, Tampa, FL 33620, USA}

\author{Robert~W.~Boyd}
\affiliation{The Institute of Optics, University of Rochester, Rochester, New York 14627, USA}
\affiliation{Department of Physics, University of Ottawa, Ottawa, ON K1N 6N5 Canada}

\author{Gerd~Leuchs}
\affiliation{Max Plank Institute for the Science of Light, Erlangen, Germany}

\author{Zhimin~Shi}
\email[]{zhiminshi@usf.edu}
\affiliation{Department of Physics, University of South Florida, Tampa, FL 33620, USA}

\date{\today}

\begin{abstract}
Free-space optical communication is a promising means to establish versatile, secure and high-bandwidth communication for many critical point-to-point applications. While the spatial modes of light offer an additional degree of freedom to increase the information capacity of an optical link, atmospheric turbulence can introduce severe distortion to the spatial modes and lead to data degradation. Here, we propose and demonstrate a vector-beam-based, turbulence-resilient communication protocol, namely spatial polarization differential phase shift keying (SPDPSK), that can encode a large number of information levels using orthogonal spatial polarization states of light. We show experimentally that the spatial polarization profiles of the vector modes are resilient to atmospheric turbulence, and therefore can reliably transmit high-dimensional information through a turbid channel without the need of any adaptive optics for beam compensation. We construct a proof-of-principle experiment with a controllable turbulence cell. Using 34 vector modes, we have measured a channel capacity of 4.84 bits per pulse (corresponding to a data error rate of 4.3\%) through a turbulent channel with a scintillation index larger than 1. Our SPDPSK protocol can also effectively transmit 4.02 bits of information per pulse using 18 vector modes through even stronger turbulence with a scintillation index of 1.54. Our study provides direct experimental evidence on how the spatial polarization profiles of vector beams are resilient to atmospheric turbulence and paves the way towards practical, high-capacity, free-space communication solutions with robust performance under harsh turbulent environments.

\end{abstract}

\maketitle


\section{Introduction}
Free-space optical communication offers flexibility, security and large signal bandwidth as compared to other means of communication \cite{khalighi_IEEE2014:survey_op_comm,roh_IEEE2014:millimeter_wave,chan_jlt2006:fso_comm}. Recently, there has been a great amount of research interest in using spatially structured light for optical communication \cite{gibson_oe2004:fso_oam, curtis_prl2003:structure_vortex,wang_pr2016:adv_comm_ovort} as the spatial modes provide a new degree of freedom to encode information, thereby greatly increasing  the system capacity and spectral efficiency within a finite spatial-bandwidth of an optical channel. Among various families of spatial modes that haven been investigated, the orbital angular momentum (OAM) modes of light have been used most widely and  successfully to increase the information capacity of a free-space optical link \cite{wang_natphoto2012:terabit_fs_oam,willner_ol2015:20Gbits_oam,yan_natcomm2014:highcap_comm_oam,willner_RSA2017:adv_oam,li_oe2018:multi_oam_cohcoup}. However, atmospheric turbulence can impose serious practical limitations on the utilization of OAM or other spatial modes, as the fluctuation in refractive index of air can alter both the transverse amplitude and phase structures of OAM modes. This, in turn, leads to crosstalk between neighboring OAM modes and degradation of the information capacity of the free-space optical link \cite{paterson_prl2005:turb_oam, cheng_2009:prop_vortex_beam_turb, tyler_ol2009:inful_turb_oam, malik_oe2012:influ_turb_oam_code, ren_ol2013:turb_fso_multiplex}. Adaptive optics has been used as the standard approach to compensate for distortions caused by thin turbulence \cite{ren_ol2014:adaptive_oam,yin_ao2018:adaptive_OAM,wang_oe2018:adaptive_waw_oam}, but it remains a challenge to compensate for beam distortions caused by strong or volumetric turbulence. Other approaches such as image recognition based on artificial intelligence and machine learning \cite{krenn_njop2014:communication_Vienna,doster_ao2017:ml_OAM_neural_net,park_oe2018:vb_oc_pattern_rocog,lohani_ol2018:turbulence_neural_net} have also been demonstrated to resolve the information encoded in severely distorted structured beams through turbulence. However, these approaches require both the acquisition of images and significant computing resources, making them not suitable for high-speed operation in real time.

Meanwhile, vector beams  \cite{Zhan_AOP2009:vector_beam_review} are optical fields that possess non-uniform spatial profiles in both complex-amplitude and polarization.  The diversity in degrees of freedom within the vectorial optical fields has brought new dimensions for fundamental studies \cite{Kagalwala_NatPhoton13:BellMeasureClassicalCoherence, Bauer_Sci15:PolarMobiusStrip, Larocque_NatPhys18:TologolyPolKnots}  and led to new optical applications  \cite{Abouraddy_PRL2006:VB_microscopy, Zhan_OE2002:focus_vector_beam, Donato_OL2012:Vb_optical_trap, Dunlop_JO2016:roadmap_struct_light, Beresna_OME2011:Polar_fab_nano_glass, otte_LSA2018:entanglement,ndagano_Natphy2017:charac_states_light, deAguiar_SciAdv17:PolRecoveryScattering,Moradi_oe2019:vector_beam_trap} with performances surpassing those of conventional approaches.  In particular, it has been shown that atmospheric turbulence is polarization-insensitive, and therefore the spatial polarization profiles of vector beams are more resilient to atmospheric turbulence as compared to the transverse phase profiles  \cite{andrews2005:laser_random,cheng_oe2009:prop_vb_turb, Cox_OE16:Resilience_VectorScalarVortexBeam}. Interestingly, many studies\cite{zhao_ol2015:high_vb_encode, milione_ol2015:multiplex_fs_vb, Cox_OE16:Resilience_VectorScalarVortexBeam,Sit_optica2017:HD_crypto_photon, ndagano_Natphy2017:charac_states_light, ndagano_2018:creation} show that vector-beam-based protocols do not outperform their scalar-beam-based counterparts, and both are equally vulnerable to atmospheric turbulence. Thus, it remains a challenge to effectively utilize a large number of spatial modes to transmit information through a turbid channel.

	In this work, we propose a new information encoding protocol, namely, spatial polarization differential phase shift keying (SPDPSK), that encodes high-dimensional information on orthogonal spatial polarization states of a family of vector vortex beams. We observe experimentally that the spatial polarization profile of vector vortex beams is resilient against atmospheric turbulence. By utilizing such advantages, our SPDPSK protocol can transmit high-dimensional information reliably through a moderately strong turbulence cell without the need of any beam compensation mechanism. We demonstrate a proof-of-principle, high-dimensional communication system by transmitting 34 information levels (5.09 bits of information) per pulse through a free-space channel in the moderately strong turbulence regime with small information loss. 
\section{Principle}

\begin{figure}
\includegraphics{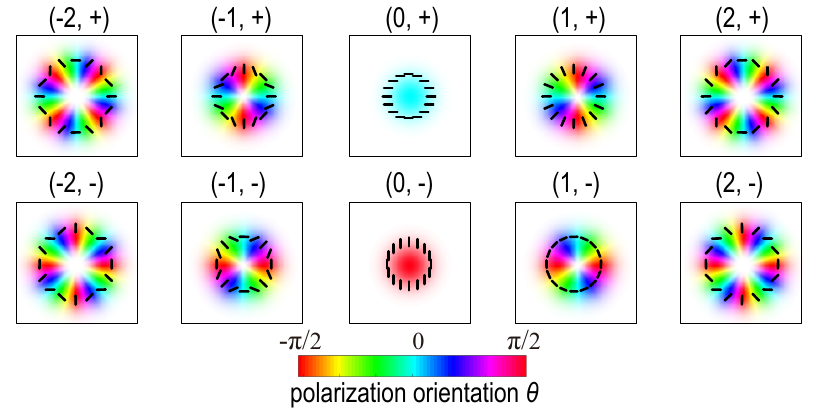}
\caption{\label{fig:VectorModes} The spatial polarization profile of 10 vector beams comprised of Laguerre-Gaussian (LG) beams in the circular polarization basis with opposite OAM charges along with $0$ or $\pi$ relative phase difference. The color indicates the polarization orientation while the color saturation indicates the beam intensity. The black lines indicate the local polarization state across the beams.}
\end{figure}

We here propose to use a family of vector vortex beams with orthogonal spatial polarization profiles to represent a large number of information levels. As an example, we consider a family of vector vortex beams, each formed by superposing two Laguerre-Gaussian (LG) beams that possess OAM charges of opposite signs in the two circular polarization bases. The $m$th order LG vector vortex beam can be expressed as follows:
\begin{eqnarray}
\label{eqn:DefVectorModes}
\vec{E}_{m,\pm}(r,\theta,z) & = &  \hat{e}_{\ell} E_{\ell, m, \pm}(r,\theta,z) + \hat{e}_{r} E_{r, m, \pm}(r,\theta,z) \nonumber \\
& = & \hat{e}_{\ell}{\rm LG}_{0,m}(r,\theta,z) \pm \hat{e}_{r}{\rm LG}_{0,-m}(r,\theta,z),
\end{eqnarray}
where ${\rm LG}_{0,m}$ denotes the spatial field profile of a Laguerre-Gaussian beam with radial index $p=0$, azimuthal index $m$, and the subscript $+$ or $-$ indicates the relative phase difference of 0 or $\pi$ between the two polarization components. In this configuration, allowing the mode order index $m$ to possess $N$ different values, we can have a total number of $2N$ orthogonal vector vortex beams to represent $2N$ information levels. Figure~\ref{fig:VectorModes} shows the spatial polarization profile of ten such vector vortex beams with mode order index, $m = -2, -1, 0, 1, 2$, and with a relative phase difference of $0$ and $\pi$ between the two circular polarization components. These 10 vector vortex beams represent 10 different information levels in our protocol.

The vector vortex beams then propagate through a free-space optical link. At the receiver end, the beam is first split into $N$ copies. Each copy then passes through a decoding channel designed for the identification of the $n$th order vector modes. In each decoding channel, the beam first passes through an anisotropic decoding phase plate with a differential phase response of $\Delta \phi_n(r, \theta) = 2 n \theta $ between the right- and left-handed circular polarization components. The decoded beam then passes through a polarizing beam splitter (PBS), and subsequently a balanced detector is used to measure the power difference between the separated H and V polarization components. When a $m$th order vector vortex beam passes through a $n$th order decoding channel, the normalized detected signal is given by
\begin{eqnarray}
P^{n}_{m, \pm}  = {\Bigg \{} \begin{array}{cc}
\pm 1 & n = m \\
0 & n \neq m \\
\end{array}.
\end{eqnarray}

\begin{figure}
\includegraphics{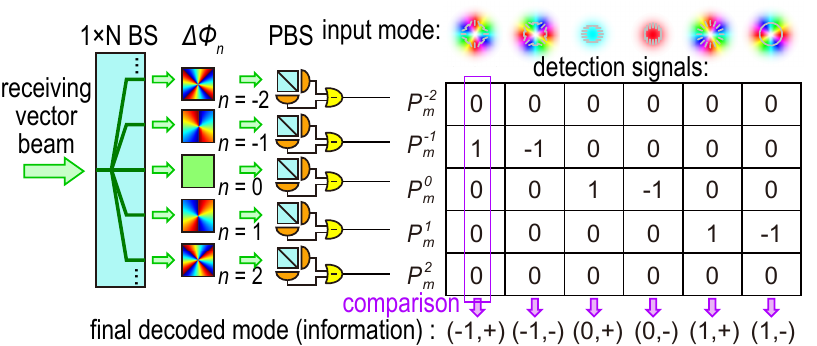}
\caption{\label{fig:DetPrinciple} The principle of signal detection. The incoming optical signal is split into $N$ copies, each passing through a $n$th order polarization-dependent, decoding phase mask before the differential power between H and V polarizations is measured. All $N$ decoded signals are then compared to determine the final detected information level (vector beam modes). The table on the right shows the detected signals of five decoding channels for six input modes (shown on the top), and the final decoded information level on the bottom.}
\end{figure}

As one sees, for the $n$th order decoding channel, an incoming $n$th order vector beam would result in detection signal of 1 or -1 for the $(n, +)$ and $(n, -)$ input modes, respectively. On the other hand, an incoming $m$th order vector vortex beam would result in a detection signal of 0 if $m \neq n$. All $N$ differently decoded detection signals for the same received beam are then compared to determine its information level. An illustration of the detection process for the six lowest order vector beams with five decoding channels is shown in Fig.~\ref{fig:DetPrinciple}. Since the information is encoded in the spatial polarization profile of the beam, which is determined by the spatially-varying phase difference between the two polarization components of the vector beam, we name our protocol {\it spatial polarization differential phase shift keying} (SPDPSK). 

When the vector beam propagates through a turbulent channel, both circular polarization components of the beam can experience severe phase and amplitude distortions. However, because the atmospheric turbulence is insensitive to the polarization, the difference of the distortion between the two polarization components, which determines the distortion of the spatial polarization profile, tends to be much smaller than the distortion of the complex-field profile of each polarization component. Since the information is encoded in the spatial polarization profile rather than the complex-field profile of the beam, it is better conserved under turbulent conditions. The turbulence resilience of our SPDPSK protocol can also be understood analogously to the well-known differential phase shift keying (DPSK) protocol being resilient to phase fluctuation in optical fibers. More detailed descriptions of the theoretical framework of our protocol are given in the supplemental material.

\section{Experimental Results and Analysis}

\begin{figure}
\includegraphics{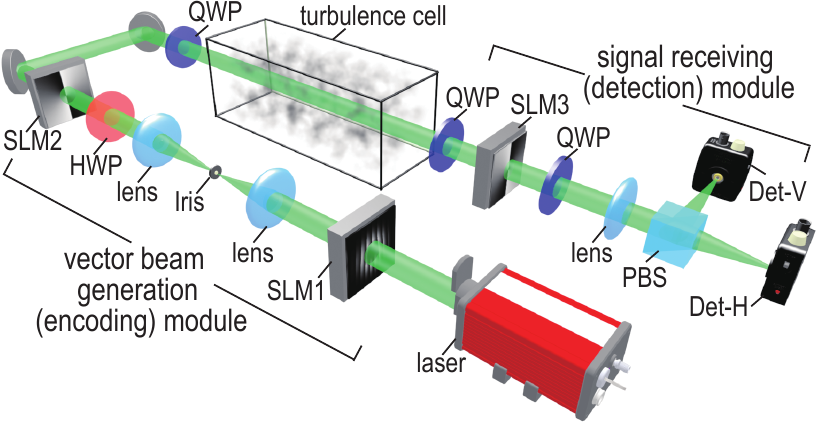}
\caption{\label{fig:Schematics} The schematic diagram of the proof-of-principle experiment including a vector beam generation module, a controllable turbulence cell and a signal detection module. SLM: spatial light modulator; HWP: half-waveplate; QWP: quater-waveplate; PBS: polarizing beam splitter;  Det: detector.}
\end{figure}

To examine the performance of our SPDPSK protocol under various atmospheric turbulence conditions, we construct a proof-of-principle experiment with a controllable turbulence cell. As illustrated in Fig.~\ref{fig:Schematics}, we use two cascaded phase-only spatial light modulators (SLMs) with polarization optics to generate the desired vector vortex beams. The generated beam is expanded and sent through a hotplate-based turbulence cell, wherein the strength of turbulence can be controlled by adjusting the hotplate temperature. At the receiving end, the beam is demagnified before propagating through one decoding channel. The beam is reflected off from a third SLM (SLM3 in Fig.~\ref{fig:Schematics}), which is used as the anisotropic decoding phase plate. The decoded beam then passes through a polarizing beam splitter (PBS), and the powers of the separated horizontal and vertical components of the decoded beam are measured by two photodetectors. The two power readings are then subtracted to give the final detection signal of the current decoding channel. The full details of the experimental setup are described in the supplemental material.

\begin{figure}
\includegraphics{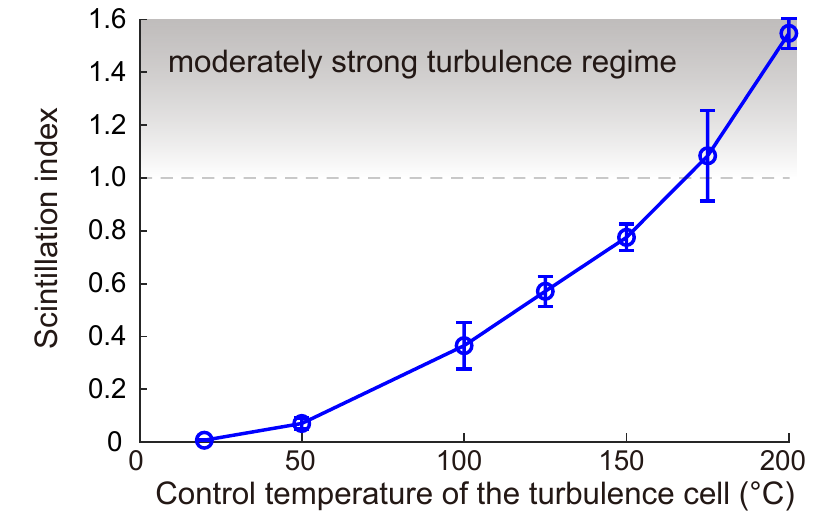}
\caption{\label{fig:Scintillation} Measured scintillation of Gaussian beams propagating through the turbulence cell as a function of the control temperature.}
\end{figure}

We first characterize the turbulence strength of our free-space channel by propagating a large Gaussian beam through the channel and measuring its scintillation index at the decoding SLM3 plane. The scintillation index is defined as follows \cite{andrews_2001:laser_scintillation, gu_ol2009:scint_polar_beam}:
\begin{eqnarray}
\sigma_I^2 = \frac{\langle I^2 \rangle }{\langle I \rangle^2} - 1,
\end{eqnarray}
where $\langle \cdot \rangle$ denotes ensemble average, and $I$ is the measured beam intensity at the statistical center of the beam. The measured scintillation index as a function of control temperature of the hotplate is shown in Fig.~\ref{fig:Scintillation}. One sees that the measured scintillation index value ranges from approximately 0 to 1.54, which indicates that our setup can provide up to moderately strong turbulence  \cite{andrews_2001:laser_scintillation} that is controllable and adjustable. 

\begin{figure*}
\includegraphics{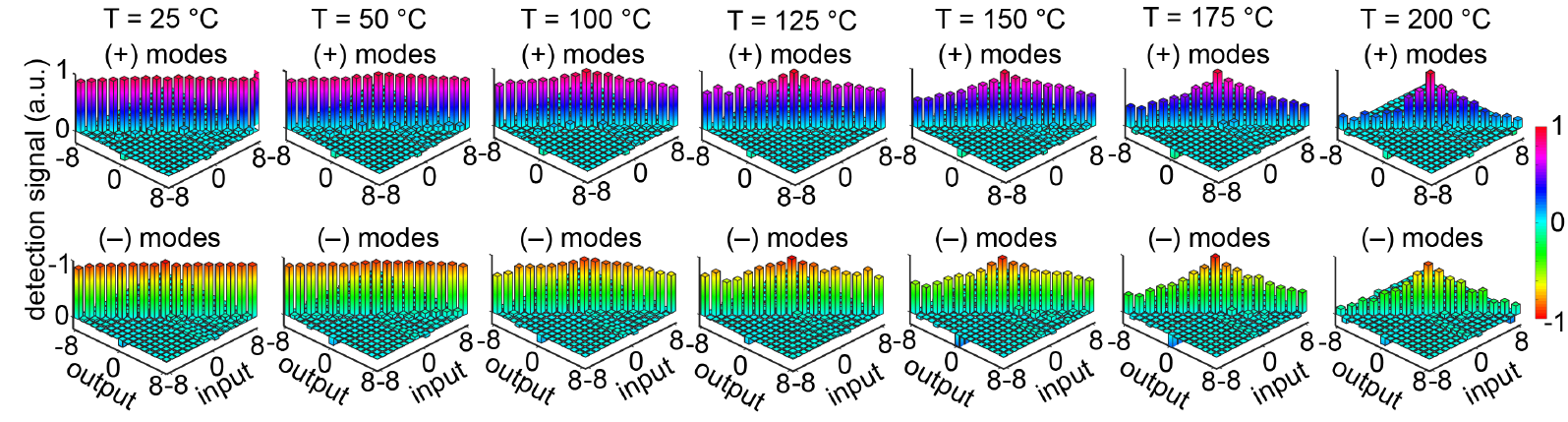}
\caption{\label{fig:AverageSignalMatrix} The average values of the directly measured signal of each input mode through each decoding (output) channel at various turbulence strengths. The top and bottom rows are for $(+)$ and $(-)$ input modes as described in Eq.~(\ref{eqn:DefVectorModes}), respectively.}
\end{figure*}

In our communication experiment, we use 17 mode orders ($m = -8$ to $8$), which enables a total of 34 information levels, i.e., 5.09 bits of information per pulse. For each transmitted mode at a given turbulence strength, we measure the detection signals through each decoding channel in sequence. A total of 2500 measurements are taken for each encoding-decoding configuration, and the averaged detection signal matrices for all the $(+)$ and $(-)$ input modes at various turbulence levels are shown in Fig.~\ref{fig:AverageSignalMatrix}. As one sees from the figure, when the channel is turbulence-free, the correctly decoded signals of all 17 mode orders (the diagonal elements) are approximately 1 (or -1) for the $(+)$ and $(-)$ modes, respectively. Meanwhile the incorrectly decoded detection signals (the off-diagonal elements) remain very small and close to zero. As the turbulence strengthens, the average values of the correctly decoded signals gradually drop, more rapidly in the case of higher mode orders. This is because higher order modes exhibit a larger difference in the wavefronts of the two polarization components, which leads to a larger difference in the distorted transverse profile as they co-propagate through the turbulence. The average values of the incorrectly decoded signals remain near zero. In general, the fluctuations of all the detected signals increase at higher turbulence strengths and for larger mode orders.

\begin{figure}
\includegraphics{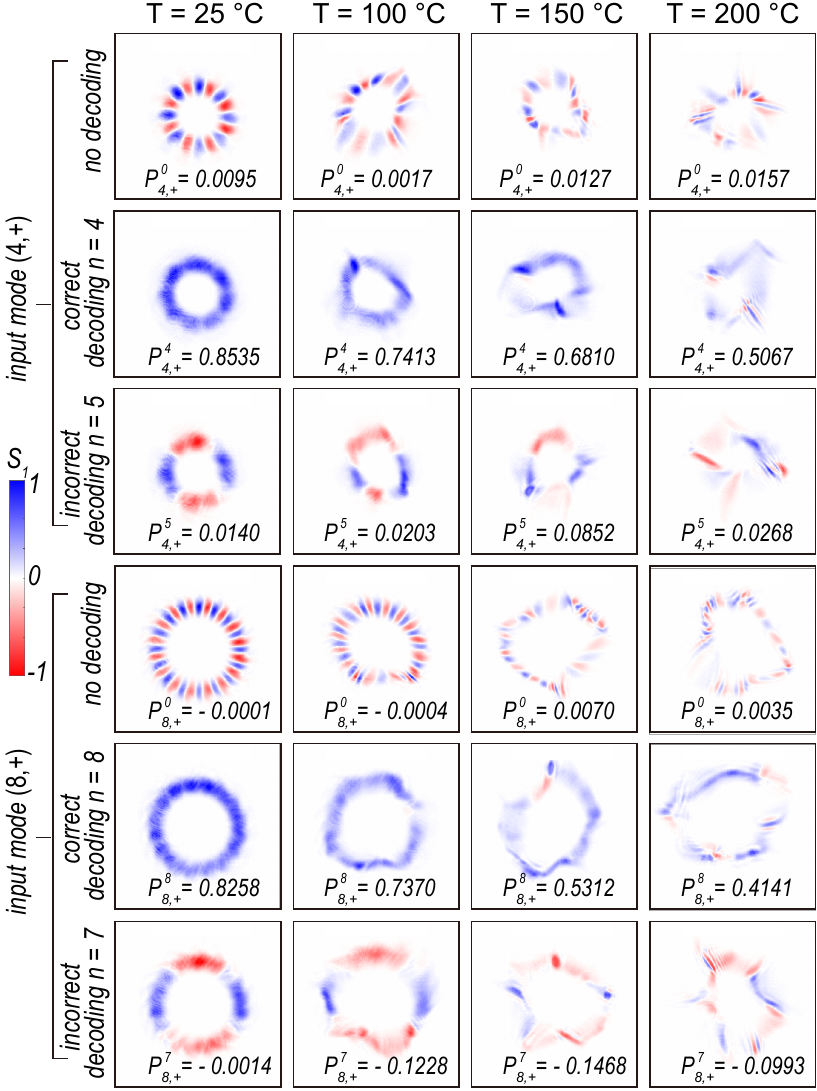}
\caption{\label{fig:S1_profile} The measured $S_1$ Stokes parameter profile of $(m=4, +)$ (upper) and $(m=8, +)$ (lower) vector vortex modes propagating through the free-space channel at various turbulence settings with no decoding, correct decoding and incorrect decoding, respectively. More snapshots of the profiles can be found in the supplemental animation clips.}
\end{figure}

To gain a more direct understanding of the preservation and eventual degradation of the spatial polarization profiles through turbulence, we use a polarization-resolving camera to capture the spatial polarization profiles of two representative vector vortex beams, $(m=4, +)$ and $(m=8,+)$, after propagating through the free-space channel with various turbulence strengths and with no decoding, correct decoding, and incorrect decoding, respectively. From the no-decoding results, one sees that while the intensity profile of the vector beams experiences greater distortion at stronger turbulence levels, the polarization profile is in fact much better preserved. As a result, when the beam is decoded correctly, the vector beam is transformed into a scalar one that can lead to a large detection signal. Note that the detection signal listed blow each $S_1$ profile in Fig.~\ref{fig:S1_profile} is calculated by summing over the $S_1$ profile directly captured by the polarization-resolving camera with a relatively low polarization extinction ratio, leading to lower detection signal values as compared to the results shown in Fig.~\ref{fig:AverageSignalMatrix} that are obtained by a PBS and two photodetectors. As the channel enters into the moderately strong turbulence regime, i.e., a scintillation index larger than 1, corresponding to the hotplate control temperature, $T>150\,^\circ$C, the polarization profiles of the transmitted vector beams experience more significant distortions resulting in the decoding becoming less effective. Furthermore, the polarization profile of higher order vector modes become more distorted, which explains the more rapid drop in their correctly-decoded detection signals as shown in Fig.~\ref{fig:AverageSignalMatrix}. On the other hand, when the beam is decoded incorrectly with a mismatching polarization phase mask, the power of the beam is still well split between the two polarizations as indicated by equally distributed blue and red colors. This explains the incorrectly decoded detection signals are all close to zero, with the largest non-zero average values occurring at $n=m \pm 1$.

\begin{figure*}
\includegraphics{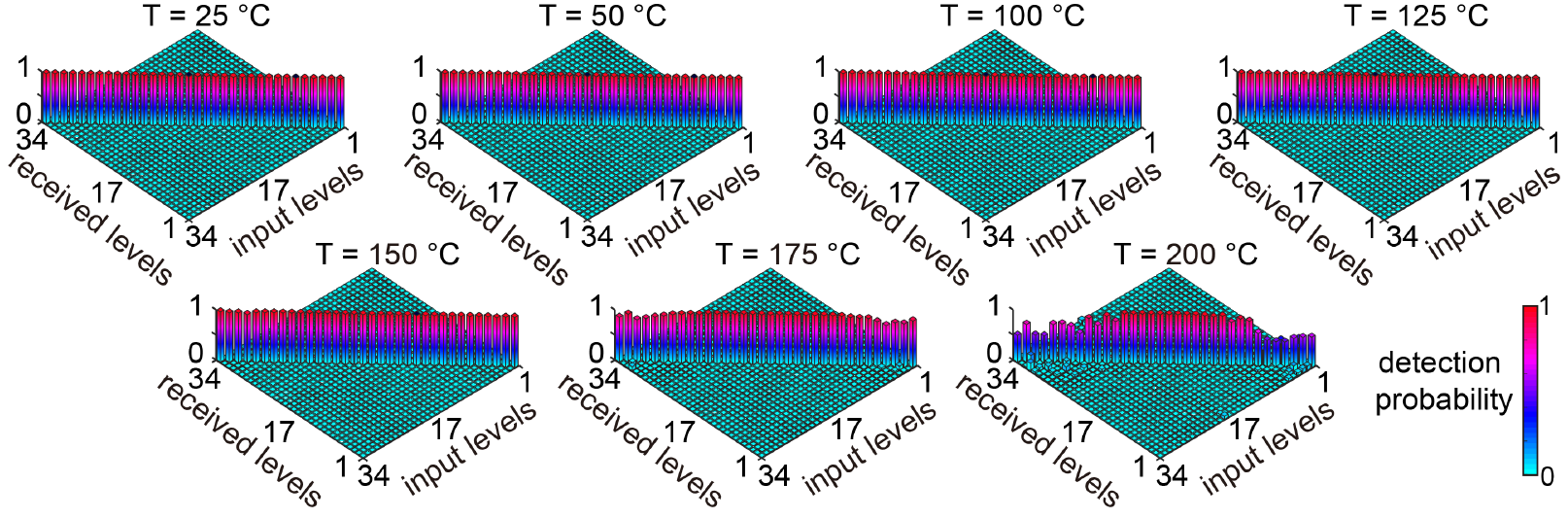}
\caption{\label{fig:DetProbMatrix} The experimentally measured detection probability matrix for the transmitted and received information levels at different turbulence strengths.  The horizontal axes denote the encoded and received information levels.}
\end{figure*}

The final information level of each received beam is determined by selecting the largest positive (or negative) value among the 17 detection signals, each through a different decoding channel. Note that in a full-scale system, all detection signals should be obtained simultaneously for each incoming beam. However, our numerical simulation indicates that the fluctuations in the detection signals of different decoding channels due to turbulence are quite uncorrelated. Thus, as a proof-of-principle demonstration, we perform the signal detection in different decoding channels sequentially and then compare them to determine the received information level. The obtained information detection probability matrices for all of the 34 information levels are shown in Fig.~\ref{fig:DetProbMatrix}. As one sees, the probability of correctly detecting the information levels remains approximately unity even when the control temperature of the turbulence cell is 175 $^\circ$C, corresponding to a scintillation index of $1.09$.


Based on the experimentally measured detection probability matrix, one can further compute the average optical signal error rate and the channel mutual information according to the following formulas:
\begin{eqnarray}
{\rm ER}(2N) & = &  \sum_{\alpha=1}^{2N} P_\alpha \sum_{\beta \neq \alpha} P_{\beta|\alpha}, \\
{\rm MI}(2N) & = &  -\sum_{\alpha=1}^{2N}  P_\alpha \log_2(P_\alpha) \nonumber \\
& & + \sum_{\alpha=1}^{2N}  P_\alpha \sum_{\beta \neq \alpha} P_{\beta|\alpha} \log_2(P_{\beta|\alpha} ),
\end{eqnarray}
where $2N$ is the total number of vector modes used, and $P_\alpha$ is the probability of information level $\alpha$ in the encoded data stream. Here we assume all levels are equally probable such that $P_\alpha = 1/(2N)$, and $P_{\beta|\alpha} $ is the conditional probability of detecting an incoming information level $\alpha$ as information level $\beta$.

\begin{figure}
\includegraphics{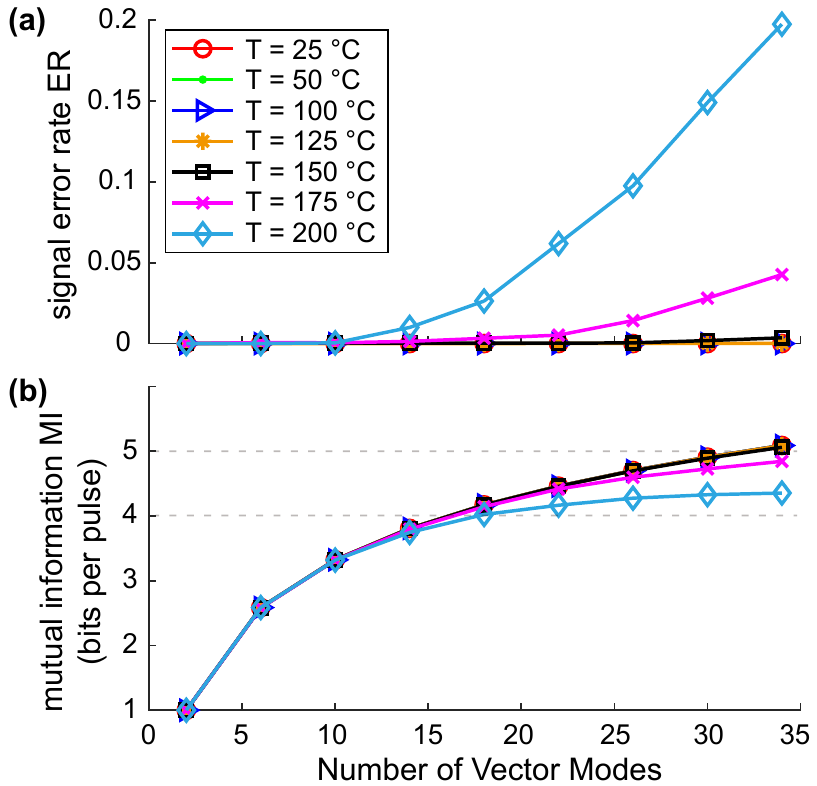}
\caption{\label{fig:ER_MI_vs_ModeNum} The experimental (a) signal error rate and (b) mutual information as functions of the number of vector modes used in the system at various temperatures. Here all levels are considered equally probable in the data stream.}
\end{figure}

Figure~\ref{fig:ER_MI_vs_ModeNum} shows the average optical signal error rate, ${\rm ER}$, and the channel mutual information, ${\rm MI}$,  as functions of the number of vector modes used in the system at various temperatures (turbulence levels).   As one sees, the system is practically error-free in weak and moderate turbulence conditions (corresponding to temperatures under 150 $^\circ$C), with an average signal error rate less than $0.35\%$. The mutual information between the sender and the receiver, which describes the channel capacity, follows the theoretical upper bound of $\log_2(N)$ bits per pulse, which corresponds to 5.09 bits per pulse for 34 vector modes. As the turbulence strengthens and the scintillation index becomes larger than one, the error rate starts to increase, especially due to the degradation in the spatial polarization profile of higher-order modes. For a system using 34 vector modes, the average error rate increases to $4.3\%$ and $19.7\%$ at temperatures of 175 $^\circ$C and 200 $^\circ$C, respectively. This corresponds to a channel mutual information of 4.84 and 4.35 bits per pulse, respectively. However, if one reduces the total number of vector modes to 18, the average error rate is reduced significantly to only $2.6\%$ at the highest temperature setting of 200 $^\circ$C with a scintillation index of 1.54. This corresponds to a channel capacity of 4.02 bits of information per pulse, only slightly lower than the theoretical upper limit of 4.17 bits-per-pulse for a 18-level system.

\begin{figure}
\includegraphics{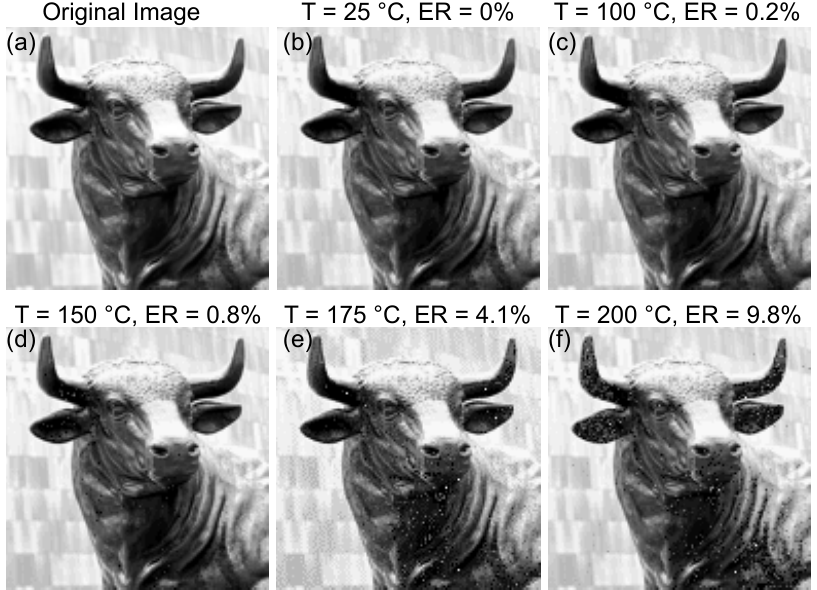}
\caption{\label{fig:Images} The retrieved 5-bit gray-scale image encoded in a 32-level SPDPSK system sent through a free-space channel with various turbulence strengths. The error rate of each image is listed on the top.}
\end{figure}

Lastly, we demonstrate the transfer of a data packet through the turbid channel using our SPDPSK communication protocol. The data packet used here is a 5-bit gray scale image with 128$\times$128 pixels. We use 32 non-zeroth-order vector modes (see supplemental material for the details of encoding table) such that the 5-bit gray scale information of each pixel is fully encoded into one pulse. The received images at various turbulence strengths are shown in Fig.~\ref{fig:Images}. One sees that our high-dimensional communication system can reliably transmit the image through a free-space channel with up to moderately strong turbulence. Additionally, the error rate measured in the image data matches well with our previous results of the detection probability. 

In summary, we have introduced a spatial polarization differential phase shift keying (SPDPSK) communication protocol that encodes high-dimensional information data onto the spatial polarization profile of an optical beam. We have shown experimentally that the spatial polarization profile of vector vortex beams can be resilient to moderately strong atmospheric turbulence, and therefore the SPDPSK protocol can transmit high-dimensional data without the need of any beam compensation mechanism. Using 34 orthogonal vector vortex beams, we have measured a channel capacity of 4.84 bits of information per pulse through a turbulent channel with a scintillation index of $1.09$. When the scintillation index was increased to $1.54$, we successfully used 18 vector modes to effectively transmit 4.02 bits of information per pulse. Our SPDPSK protocol provides a practical and robust solution for high-capacity, free-space communication under natural, harsh environments. 



Z.~Z., M.~J.,  A.~F., D.~H., Y.~Z., B.~K., T.~W., R.~W.~B and Z.~S. acknowledge support from ONR grant N00014-17-1-2443. R.W.B. acknowledeges funding from the Natural Sciences and Engineering Research Council of Canada and the Canada Research Chairs program.

\section{Supplemental Material}
\subsection{\label{sec:Theory}Theoretical Framework}

We here describe in detail the theoretical framework of our SPDPSK protocol. As an example, each vector vortex beam can be formed by a superposition of two Laguerre-Gaussian (LG) beams that possess OAM charges of opposite signs in the two circular polarization bases along with a relative phase difference of 0 or $\pi$. Such a LG vector vortex beam can be expressed as follows:
\begin{eqnarray}
\hspace{-0.4 cm} \vec{E}_{m,\pm}(r,\theta,z) & = &  \hat{e}_{\ell} E_{\ell, m,\pm}(r,\theta,z) + \hat{e}_{r} E_{r , m,\pm}(r,\theta,z) \nonumber \\
& = & \hat{e}_{\ell}{\rm LG}_{0,m}(r,\theta,z) \pm \hat{e}_{r}{\rm LG}_{0,-m}(r,\theta,z),
\end{eqnarray}
where ${\rm LG}_{p,l}$ denotes the spatial field profile of a Laguerre-Gaussian beam
\begin{eqnarray}
\hspace{-0.3 cm} {\rm LG}_{p,l}(r,\theta,z)&=&\sqrt{\frac{2p!}{\pi(p+|l|)!}}\frac{1}{w(z)}\left(\frac{\sqrt{2}r}{w(z)}\right)^{|l|}e^{-\frac{r^2}{w(z)^2}}\nonumber \\
&& L^{|l|}_{p} \left(\frac{2r^2}{w(z)^2}\right)e^{-ik\frac{r^2}{2R(z)}}e^{il\theta}e^{-ikz}e^{i\phi(z)},
\end{eqnarray}
and where $p$ and $l$ denote the radial and azimuthal index, respectively, $L^{|l|}_{p} ()$ is the generalized Laguerre polynomial, $w(z)=w_0\sqrt{1+(z/z_R)^2}$ is the beam waist, $z_R = \pi w_0^2/\lambda$ is the Rayleigh range and $R(z)=z[1+(z_R/z)^2]$ is the radius of curvature of the beam. If $m$ takes $N$ different values, we have a total number of $2N$ orthogonal vector vortex beams to represent $2N$ information levels. 

Since the spatial polarization profile is determined by the relative phase (and amplitude) of the two orthogonal polarization components, the information encoded in the spatial polarization profiles is essentially encoded as the difference of the spatially-varying phase profiles between the two polarization components. Thus, analogously to the well-known differential phase shift keying (DPSK) protocol in which the information is encoded in the relative phase between neighboring pulses in the time domain, we name our protocol spatial polarization differential phase shift keying (SPDPSK), indicating that the information is encoded in the relative phase between the two polarization components of a vector beam. In theory, the spatial polarization profiles can span a Hilbert space of infinitely large dimensions, indicating the number of information levels that can be encoded is infinitely large. In practice, the number of usable information levels is determined by the spatial-bandwidth product (Fresnel number) of the free-space optical link as well as the turbulence strength.


When the free-space optical channel is turbulence-free, the two polarization components of each vector vortex beam would pick up the same phase as the beam propagates through the channel; therefore the spatial polarization profile of the beam remains invariant upon propagation. Consequently, the spatially dependent phase difference between the two polarization components is  independent of the propagation distance, which can be shown as follows:
\begin{eqnarray}
\Delta \phi_{m,\pm} (r, \theta, z) & =&   {\rm arg} \left[ E_{\ell} (r,\theta,z)\right]- {\rm arg} \left[E_{r} (r,\theta,z)\right] \nonumber\\
& = & 2 m \theta + (0.5 \mp 0.5) \pi.
\end{eqnarray}

In order to decode the information at the receiver end, we first split the received vector vortex beam into N equal copies (see Fig.~\ref{fig:DetPrinciple}). Each copy then passes through a decoding channel designed for the identification of the $n$th order vector modes. In each decoding channel, the beam first passes through an anisotropic decoding phase plate that has a polarization-dependent transmission function. The transmission function of the $n$th-order decoding phase plate can be written in the left- and right-handed circular polarization (LCP and RCP, respectively) bases as follows:
\begin{eqnarray}
{\bf T}_{n}(r, \theta) = e^{i\phi_{\ell}(r, \theta)}  \left[ \begin{array}{cc}
1 &   0 \\
0   &  e^{i \Delta \phi_n (r, \theta) } \\
\end{array}\right],
\end{eqnarray}
where $e^{i\phi_{\ell}(r, \theta)}$ is some transmission function for the left-handed circularly polarized light, and $\Delta \phi_n(r, \theta) = 2 n \theta $ is the difference between the phase responses for the right- and left-handed circular polarization components. When a $m$th order vector vortex beam passes through such a $n$th order decoding phase plate, the transmitted field becomes
\begin{eqnarray}
\vec{E}^{n, {\rm out}}_{m, \pm} &=& {\bf T}_{n}\vec{E}^{\rm in}_{m,\pm} \nonumber \\
&=&\hat{e}_{\ell}{\rm LG}_{0,m}e^{i\phi_{\ell}} \pm \hat{e}_{r}{\rm LG}_{0,-m}e^{i\phi_{\ell}} e^{i2n\theta}.
\end{eqnarray}
When the output beam passes through a polarizing beam splitter (PBS), the intensity of the field exiting the two output ports becomes
\begin{eqnarray}
I^{n,{\rm h}}_{m, \pm}= |{\rm LG}_{0,m}|^2 \cos^2[(m-n)\theta-(1\pm 1)\frac{\pi}{4}], \nonumber \\
I^{n,{\rm v}}_{m, \pm} = |{\rm LG}_{0,m}|^2 \sin^2[(m-n)\theta-(1\pm 1)\frac{\pi}{4}].
\end{eqnarray}
The normalized power of the two separated H- and V- components are given as follows:
\begin{eqnarray}
P^{n,{\rm h}}_{m, \pm} & = & {\Bigg \{} \begin{array}{cc}
\frac{1}{2} \pm \frac{1}{2} & n = m \\
0.5 & n \neq m \\
\end{array}, \nonumber \\
P^{n,{\rm v}}_{m, \pm} & = & {\Bigg \{} \begin{array}{cc}
\frac{1}{2} \mp \frac{1}{2} & n = m \\
0.5 & n \neq m \\
\end{array}.
\end{eqnarray}

The final $n$th order detection signal is obtained by taking the difference between the two above outputs, which is given by
\begin{eqnarray}
P^{n}_{m, \pm} \equiv P^{n,{\rm h}}_{m, \pm} - P^{n,{\rm v}}_{m, \pm} = {\Bigg \{} \begin{array}{cc}
\pm 1 & n = m \\
0 & n \neq m \\
\end{array}.
\end{eqnarray}

As one sees, for the $n$th order decoding channel, an incoming $n$th order vector beam would result in a detection signal of 1 or -1 depending upon the relative phase between the LCP and RCP components. On the other hand, an incoming $m$th order vector vortex beam would result in a detection signal of 0 if $m \neq n$ in the absence of turbulence.


\subsection{Experimental Procedures}
To demonstrate the performance of our proposed SPDPSK protocol, we construct a proof-of-principle experiment in a laboratory setting. As shown in Fig.~\ref{fig:Schematics}, our experimental setup is comprised of a vector beam generation module, a controllable turbulence cell, and a signal detection module. To generate the desired $m$th order vector vortex beam, a laser beam from a 532-nm laser (Coherent Compass M315) with horizontal polarization is first expanded, collimated and launched onto a phase-only spatial light modulator (SLM1; CambridgeCorrelaters SDE1024). A computer-generated hologram (CGH) for ${\rm LG}_{0,m}$ is imprinted onto SLM1 \cite{Davis_AO99:AmpModwphaseSLM}, and the diffracted light passes through a 4-$f$ imaging system with spatial filtering at the focal plane. The transmitted light in the first diffraction order is adjusted to be 45 degree polarized before it reaches a second phase-only SLM (SLM2; Hamamatsu) placed at the image plane, which is also responsive only to horizontal polarization. SLM2 is imposed with a phase profile of $-2m\theta$ or $-2m\theta  + \pi$ for the $(+)$ and $(-)$ modes, respectively. As a result, the horizontal and vertical components of the resultant beam share identical amplitude profiles of a Laguerre-Gaussian mode but with opposite OAM charges and an overall relative phase shift of $0$ or $\pi$. The beam further passes through a quarter-wave plate to convert the H and V polarizations into left- and right-handed circular polarizations, respectively, to become the desired $m$th order vector vortex beams, $\vec{E}_{m, \pm}$.

The generated vector vortex beam is then expanded by a 3.3x telescope before it propagates through a hotplate-based turbulence cell approximately 60 cm in length. The strength of the turbulence is controlled by adjusting the temperature of the hotplate. The transmitted beam then passes through a receiving telescope with 3.3x demagnification and a quarter-waveplate before being launched onto a decoding spatial light modulator (SLM3; Hamamatsu). The decoding SLM3 responds only to horizontal polarization, and is imposed with a spiral phase profile of $\Delta \phi_n = 2n\theta$ for the $n$th-order decoding channel. The decoded beam then passes through a quarter-waveplate and a polarizing beam splitter, and the two outputs are focused onto two photodetectors.

Note that in a full-scale system, the vector beam that reaches the receiving end should first pass through a $1-{\rm to}-N$ beam splitter, resulting in $N$ identical copies, where $N$ is the total order number of vector modes used. Each copy would then pass through a decoding module designed for the identification of the $n$-th order vector mode, and a total of $N$ final detection signals would be obtained simultaneously for each incoming pulse.  The information (spatial polarization mode) of the received signal is then determined by comparing these $N$ signals. However, our numerical simulation indicates that the fluctuations in these $N$ signals for any input mode due to turbulence can be considered uncorrelated. Therefore, our experimental demonstration of collecting the detecion singal of one decoding channel at a time is still valid in assessing the data fidelity and error rate with a similar performance to a full-scale system. 

\subsection{Image Data Transfer}
\begin{figure}
\includegraphics{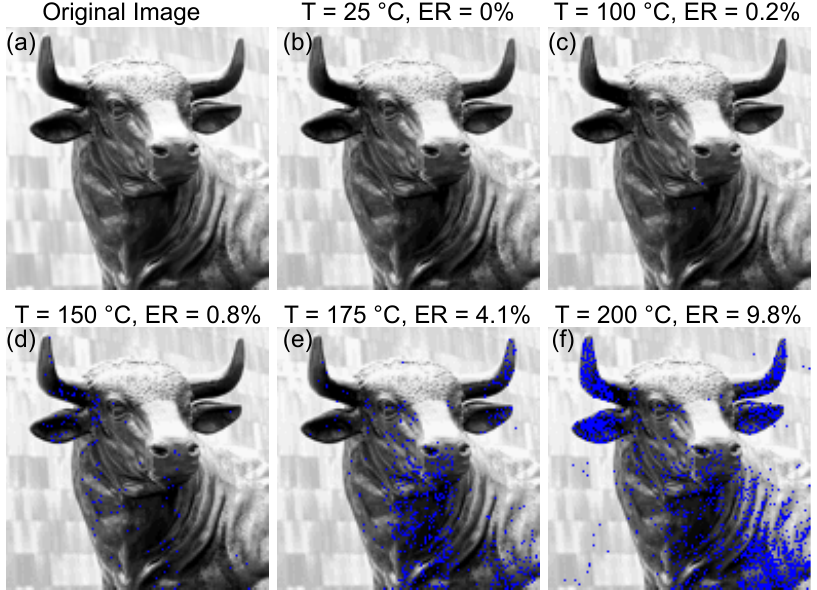}
\caption{\label{fig:Images_error} The retrieved 5-bit gray-scale image encoded in a 32-level SPDPSK system sent through a free-space channel with various turbulence strengths. The blue pixels indicate the incorrect data received.}
\end{figure}

To represent the 5-bit gray-scale information of each pixel in the picture, we use 32 vector vortex beams with order index, $m = -8, -7, ... -1, 1, ... 7, 8$. The relation between each gray scale and the corresponding vector vortex mode index is listed in table~\ref{tab:image_LUT}. Figure.~\ref{fig:Images_error} shows the received images at various turbulence levels, in which the pixels whose gray level information is received incorrectly are marked in blue.  It can be seen that our high-dimensional communication system can reliably transmit the image through a free-space channel with up to moderately strong turbulence, and that the error rate measured in the image data correlates well with our previous results of the detection probability. 

\begin{table}[]
\caption{5-bit data encoding look-up table}
\label{tab:image_LUT}
\begin{tabular}{|C{2cm}|C{1cm}|C{1cm}|C{1cm}|C{1cm}|C{1cm}|C{1cm}|}
\hline & \\[-2ex]
 gray level &  0 &  1 &  2 & 3  & 4 & 5 \\[1ex] \hline & \\[-2ex]
 mode index & (8,+)  & (-8,+) &(8,-)  &(-8,-) & (7, +) & (-7, +) \\[1ex] \hline & \\[-2ex]
gray level  & 6 & 7 & 8 & 9  & 10  & 11 \\[1ex] \hline & \\[-2ex]
 mode index  & (7, -) & (-7, -)  & (6, +)  & (-6, +) & (6, -) & (-6, -)\\[1ex] \hline & \\[-2ex]
gray level  & 12 & 13 & 14 & 15 & 16 & 17 \\[1ex] \hline & \\[-2ex]
 mode index  & (5, +) & (-5, +)  & (5, -)  & (-5, -) & (4, +) & (-4, +) \\[1ex] \hline & \\[-2ex]
gray level  & 18 & 19 & 20 & 21 & 22 & 23 \\[1ex] \hline & \\[-2ex]
 mode index  & (4, -) & (-4, -)  & (3, +)  & (-3, +) & (3, -)  & (-3, -) \\[1ex] \hline & \\[-2ex]
gray level  & 24 & 25 & 26 & 27 & 28 & 29 \\[1ex] \hline & \\[-2ex]
 mode index  & (2, +) & (-2, +)  & (2, -)  & (-2, -) & (1, +)  & (-1, +) \\[1ex] \hline & \\[-2ex]
gray level  & 30 & 31 &  &  &  &  \\[1ex] \hline & \\[-2ex]
 mode index  & (1, -) & (-1, -)  &   &  &   &  \\[1ex] \hline 
\end{tabular}
\end{table}

\bigskip

\end{document}